\documentclass[aps,amsfonts,twocolumn,nofootinbib]{revtex4}
\usepackage{amssymb,axodraw}
\usepackage{latexsym,epsfig,graphicx}
\usepackage{latexsym}

\usepackage{axodraw}

\def\de{\partial}
\def\a{\alpha}
\def\b{\beta}

\def\d{\delta}

\def\la{\lambda}
\def\La{\Lambda}
\def\k{\kappa}
\def\m{\mu}
\def\n{\nu}
\def\r{\rho}

\def\s{\sigma}

\def\th{\theta}

\def\de{\partial}
\def\a{\alpha}
\def\b{\beta}

\def\d{\delta}

\def\l{\lambda}
\def\k{\kappa}
\def\m{\mu}
\def\n{\nu}
\def\r{\rho}

\def\s{\sigma}

\def\th{\theta}

\newcommand{\be}{\begin{equation}}
\newcommand{\ee}{\end{equation}}
\newcommand{\bea}{\begin{eqnarray}}
\newcommand{\eea}{\end{eqnarray}}
\newcommand{\beqar}{\begin{eqnarray*}}
\newcommand{\eeqar}{\end{eqnarray*}}

\newcommand{\ie}{{\it i.e.,}\ }
\newcommand{\reef}[1]{(\ref{#1})}

\newcommand{\nn}{\nonumber}

\def\ie{\hbox{\it i.e.}{}}

\catcode`\@=11

\@addtoreset{equation}{section}

\def\@normalsize{\@setsize\normalsize{15pt}\xiipt\@xiipt
\abovedisplayskip 14pt plus3pt minus3pt%
\belowdisplayskip \abovedisplayskip
\abovedisplayshortskip  \z@ plus3pt%
\belowdisplayshortskip  7pt plus3.5pt minus0pt}
\def\small{\@setsize\small{13.6pt}\xipt\@xipt
\abovedisplayskip 13pt plus3pt minus3pt%
\belowdisplayskip \abovedisplayskip
\abovedisplayshortskip  \z@ plus3pt%
\belowdisplayshortskip  7pt plus3.5pt minus0pt
\def\@listi{\parsep 4.5pt plus 2pt minus 1pt
            \itemsep \parsep
            \topsep 9pt plus 3pt minus 3pt}}

\def\underline#1{\relax\ifmmode\@@underline#1\else
        $\@@underline{\hbox{#1}}$\relax\fi}
\@twosidetrue \relax

\catcode`@=12

\catcode`\@=11

\def\thesection{\arabic{section}.}
\def\thesubsection{\arabic{section}-\arabic{subsection}.}
\def\thesubsubsection{\arabic{section}-\arabic{subsection}-\arabic{subsubsection}.}

\def\ps@headings{\def\@oddfoot{}\def\@evenfoot{}
\def\@oddhead{\hbox{}\hfill
        \makebox[.5\textwidth]{\raggedright\ignorespaces --\thepage{}--
        \hfill }}
\def\@evenhead{\@oddhead}
\def\subsectionmark##1{\markboth{##1}{}}
\def\subsubsectionmark##1{\markboth{##1}{}}
}

\ps@headings

\catcode`\@=12

\begin{document}


\title{Cosmology in Six Dimensions\footnote{Plenary talk at 100 Years of Relativity: International
Conference on Classical and Quantum Aspects of Gravity and
Cosmology, Sao Paulo, Brazil,  22-24 August 2005. \\
\underline{~~~~~~~~~~~~~~~~~~~~~~~~~~~~~~~~~~~~~~~~~~~~~~~~~}\\
e-mail address: lpapa@central.ntua.gr \\}}

\author{Eleftherios Papantonopoulos}

\affiliation{Department of Physics, National Technical University
of Athens,\\ GR~157~73~Athens, Greece}


%
%


%

%
%

%





\begin{abstract}

We discuss cosmological models in six-dimensional spacetime. For
codimension-1 branes, we consider a (4+1) braneworld model and
discuss its cosmological evolution. For codimension-2 branes, we
consider an infinitely thin  conical braneword model in the
presence of an induced gravity term on the brane and a
Gauss-Bonnet term in the bulk. We discuss the cosmological
evolution of isotropic and anisotropic matter on the brane. We
also briefly discuss cosmological models in six-dimensional
supergravity.

\end{abstract}
\maketitle




\def\thesection{\arabic{section}}

\section{Introduction}

\def\thesection{\arabic{section}.}

Recently, there have been many observational and theoretical
motivations for the study of theories with extra spacetime
dimensions and in particular the braneworld scenario. From the
observational side, the current paradigm, supported
 by many recent observations like the cosmic microwave background
anisotropies~\cite{Spergel:2003cb}, large scale galaxy
surveys~\cite{Scranton:2003in} and type IA
 supernovae~\cite{Riess:1999ka,Goldhaber:2001ux} suggest that
 most of the energy content  of our universe is in the form of dark matter and dark
energy. Although there have been many plausible explanations for
these dark components, it is challenging
 to try to explain these exotic ingredients of the universe using alternative gravity
 theories as such of the braneworlds. From the theoretical side, such extra-dimensional
 braneworld models are ubiquitous in theories like string or M-theory. Since
 these theories claim to give us a fundamental description of nature, it is important to study
what kind of gravity dynamics they predict.  The hope is to
propose such modified gravity theories, which share many common
features with general relativity, but
 at the same time give alternative non-conventional cosmology.

The essence of the braneworld scenario is that the Standard Model,
with its matter and gauge interactions, is localized on a
three-dimensional hypersurface (called brane) in a
higher-dimensional spacetime. Gravity propagates in all spacetime
 (called bulk) and thus connects the Standard Model sector with the internal space
dynamics. This idea, although quite old \cite{earlybranes}, gained
momentum
 the last years \cite{AADD,randall} because of its connection with string theory (for a review
 on braneworld dynamics see~\cite{Maartens:2003tw}).

Cosmology in theories with branes
 embedded in  extra dimensions has been the subject of intense investigation during the
  last years. The most detailed analysis has been done for braneworld models in five-dimensional
   space \cite{5d}. The effect of the extra dimension can modify the cosmological evolution, depending on the model,
    both at early  and late times. The
 cosmology of this and other related models
with one transverse to the brane extra dimension (codimension-1
brane models) is well
 understood (for a review see \cite{reviews}). In the cosmological generalization of \cite{randall},
 the early times (high energy limit) cosmological evolution is modified
by the square of the matter density on the brane, while the bulk
leaves its imprints on the brane by the ``dark radiation" term
 \cite{5d,RScosmology}. The presence of a bulk cosmological constant
 in \cite{randall} gives conventional
cosmology at late times (low energy limit) \cite{RScosmology}. The
early time modification  for example, can be interesting
phenomenologically because  may require less fine-tuned
inflationary parameters~\cite{Maartens:1999hf}.

In the above models, there are strong theoretical arguments for
including in the gravitational action extra curvature terms apart
from the higher dimensional
 Einstein-Hilbert term. The localized matter fields on the brane, which couple
to bulk gravitons, can generate via quantum loops  a localized
four-dimensional
 kinetic term for gravitons~\cite{oldinduced}. The latter comes in the gravitational
action as a four-dimensional scalar curvature term localized at
the position of the brane
   (induced gravity)~\cite{induced} \footnote{There has been
   a lot of discussion about the potential problem of the extra polarization states
   of the massive  gravitons regarding phenomenology (discontinuity problem),
   but the particular model seems to be consistent in a non-trivial way \cite{indconsistency}.}.
   In addition, curvature square terms in the bulk, in the  Gauss-Bonnet
 combination,  give the most general action with second-order field equations in five
dimensions~\cite{Lovelock:1971yv}. This correction is also
motivated by string theory, where
  the Gauss-Bonnet term corresponds to
the leading order quantum correction to gravity, and its presence
guarantees a ghost-free action~\cite{GB}. Let us note, however,
that if the curvature squared terms are to play an important role
in the low energy dynamics, one may face a difficulty when
interpreting gravity as an effective field theory (in the sense
that even higher dimensional operators would seem to be also
relevant).

If curvature corrections are included in \cite{randall}, the
presence of the 3-brane gives similar modifications to the
standard cosmology. In the case of a pure Gauss-Bonnet term in the
bulk, the early time cosmology is modified by a term proportional
to the matter density on the brane to the power two thirds
 \cite{GBcorrection}. If an induced gravity term
is included, the conventional cosmology is modified by the square
root of the matter density on the brane at low energies
\cite{indcosmology}. Finally, if both curvature corrections are
present, at early time the effect of both the Gauss-Bonnet and the
induced gravity terms make the universe to start with a finite
temperature \cite{idg+gb} while at late times the effect of
induced gravity make the universe to accelerate without the need
of dark energy~\cite{Deffayet}.

Braneworld models can also be extended in higher than
five-dimensions. We can consider a $(n+1)$ brane embedded in
$(n+2)$ spacetime or a $(3+1)$ brane embedded in $(n+1)$
spacetime. In this talk we will discuss $(4+1)$ cosmological
braneworld models embedded in six-dimensional spacetime
(codimension-1 models) and (3+1) cosmological braneworld models
embedded in six-dimensional spacetime (codimension-2 models).

Six or higher-dimensional braneworld models of codimension-1 are
considered as generalizations of the Randall-Sundrum model.
In~\cite{Kanti:2001vb} static and non-static solutions in a
six-dimensional bulk as well as the stability of the radion field
were discussed, while in~\cite{Arkani-Hamed:1999gq} the evolution
of the extra dimensions transverse to the brane in a Kasner-like
metric was studied.

In braneworld scenarios, contrary to the Kaluza-Klein theories,
the extra dimensions can be large if the geometry is non trivial.
If the hierarchy problem is addressed in the braneworld scenarios
for example, the extra dimensions should be large~\cite{AADD}. In
a cosmological context however, these extra dimensions are
observationally much smaller than the size of our perceived
universe. Initially the universe could have started with the sizes
of all dimensions at the Planck length. Then, a successful
cosmological model should accommodate in a natural way a mechanism
by which the extra dimensions remained comparatively small during
cosmological evolution.

Such a mechanism was proposed
in~\cite{Brandenberger:1988aj,Brand-talk}. The basic idea is that
strings dominate the dynamics of the early universe and can see
each other most efficiently in 2(p+1) (p=1 for strings)
dimensions. Therefore, strings can only interact in three spatial
dimensions, while strings moving in higher dimensions eventually
cease to interact efficiently and their winding modes will prevent
them from further expanding. If branes are included, it was shown
in~\cite{Alexander:2000xv} that strings will still dominate the
evolution of the universe at late times so the mechanism
of~\cite{Brandenberger:1988aj} still survives.

In six dimensions and for codimension-2  braneworlds, the gravity
dynamics appear even more radical and still a good understanding
of cosmology and more generally gravity in such theories is
missing. The most attractive feature of codimension-2 braneworlds
is that the vacuum  energy (tension) of the brane
 instead of curving the brane world-volume, merely induces a deficit angle in the
 bulk solution around the brane \cite{Chen:2000at} (see also \cite{stingdefects}
  for string-like defects in six dimensions). This looks very promising in relation to the cosmological
constant problem, although in existing models, nearby curved
solutions cannot be
 excluded \cite{6d}, unless one allows for singularities more severe than conical
  in particular supersymmetric models \cite{6dsusy}.  It was soon realized \cite{Cline:2003ak}
   that one can only find nonsingular
solutions if the brane energy momentum tensor is proportional to
its induced metric, which means simply that it is pure tension. A
non-trivial energy momentum tensor on the brane  causes
singularities in the metric around the braneworld which
necessitates the introduction of a cut-off (brane thickness)
\cite{Kanno:2004nr,Vinet:2004bk,Navarro:2004di}.

An alternative approach to study the gravitational dynamics of
matter on infinitely thin branes is to modify the gravitational
action as discussed previously. Indeed, it was shown in
\cite{Bostock:2003cv} that  the inclusion of a Gauss-Bonnet term
in the gravitational  action allows a non-trivial energy momentum
tensor on the brane, and in the thin brane limit,  four
dimensional gravity is recovered as the dynamics of the induced
metric on the brane. The peculiar characteristic of this way to
obtain four dimensional gravity for codimension-2 branes, is that,
apart from the inclusion of a (deficit angle independent)
cosmological constant term, there appear to be no corrections to
the Einstein equations coming from the extra dimensions in the
purely conical case. Another possibility, discussed in
\cite{intersections}, is to study (instead of conical 3-branes)
codimension-2 branes sitting at the intersection of codimension-1
branes  in the presence again of a bulk Gauss-Bonnet term.

Much less has been done, however, for the cosmology of theories in
six or higher dimensions with branes of codimension greater than
one. This is because, unlike the codimension one case, these
branes exhibit bulk curvature singularities which are worse than
$\d$-function singularities. They then need some regularization
(introduction of brane thickness) which makes the study of
cosmology on them rather complicated \cite{thickcosmo}. An
alternative way to study cosmologies of branes of higher
codimension would be  to consider corrections to the gravitational
action, such as an induced curvature term on the brane
\cite{induced} and a Gauss-Bonnet term in the bulk \cite{GB},
which allow the brane to have a mild singularity structure (see
also \cite{intersecting}). These thin brane cosmologies would have
the additional advantage that the internal structure of the brane
does not influence the macroscopic cosmological evolution.

Six-dimensional models can also be obtained in supergravity
theories. In six-dimensional supergravity theory there is a vacuum
solution which has the structure of Minkowski$\times
S^{2}$~\cite{Salam:1984cj,Randjbar-Daemi:wc}. Cosmological
applications of this solution was presented
in~\cite{Maeda:1984gq}. These theories are interesting because
many of the parameters of the models are fixed from the theory and
in particular the form of the scalar potential is dictated from
the structure of the supergravity theory. The drawbacks of these
models however is that they are plagued from anomalies. The reason
is that they are chiral and therefore a gauging of some symmetries
is required for cancellation of these
anomalies~\cite{Randjbar-Daemi:wc}.

In this talk we will discuss braneworld models of codimension-1
and codimension-2 in six dimensions analyzing their cosmology. We
will also briefly discuss cosmological models in six-dimensional
 supergravity. First we will discuss a $(4+1)$ braneworld
model in a six-dimensional spacetime bulk with a cosmological
constant. We will consider a 4-brane fixed at some position in the
sixth dimension and we derive the dynamical six-dimensional bulk
equations in normal gaussian coordinates with a cosmological
constant in the bulk and considering matter on the
brane~\cite{Abdalla:2002ir,Cuadros-Melgar:2003zh}. We look for
time dependent solutions allowing for two scale factors, the usual
scale factor $a(t)$ of the three dimensional space and a scale
factor $b(t)$ for the extra fourth dimension. If $a(t)=b(t)$ we
get the Friedmann equation of the generalized Randall-Sundrum
model in six dimensions describing a four-dimensional universe. If
$a(t) \neq b(t)$ we get a generalized Friedmann equation in six
dimensions.

The problem can be looked at a different angle
\cite{Cuadros-Melgar}. If $a=b$, we can write the bulk
six-dimensional metric in ``Schwarzschild" coordinates and then we
have the equivalent description of a 4-brane moving in a static
Schwarzschild-(A)dS six-dimensional bulk. If however $a \neq b$, a
brane observer uses $a$ and $b$, for whom they are static
quantities, to measure the departure from six-dimensional
spherical symmetry of the bulk. The important result of this
consideration is that~\cite{Cuadros-Melgar:2005ex}, since the
brane observer needs to define a cosmic time in order to derive an
effective Friedmann equation on the brane, $a$ and $b$ are related
through the Darmois-Israel junctions conditions and because of
that, their relation depends on the energy-matter content of the
brane. The physical reason of the existence of such a relation
between $a$ and $b$ is that the requirement of having a
cosmological evolution on the brane introduces a kind of
compactification on it and the relation between $a$ and $b$ acts
as a constraint of the brane motion in the bulk.

Using the six-dimensional generalized Friedmann equation and
assuming that $p \neq \hat{p} $, where $p$ is the pressure of the
physical three dimensions and $\hat{p}$ corresponds to the fourth
dimension, we make a systematic numerical study of the
cosmological evolution of the scale factors $a(t)$ and $b(t)$ for
several values of the parameters of the model, $\Lambda_{6}$ the
six-dimensional cosmological constant, $k$ the brane spatial
curvature and $w$ and $ \hat{w}$ parameterizing the form of the
brane energy-matter content of the three dimensions and of the
extra fourth dimension respectively. We find
that~\cite{Cuadros-Melgar:2005ex}, in order the fourth dimension
to be small relatively to the other three dimensions and to remain
constant during cosmological evolution, $ \hat{w}$ must be
negative, indicating the presence of a dark form of energy in the
extra fourth dimension. We find this result for all cases
considered, (A)dS or Minkowski bulk, open, closed or flat
universe, radiation, dust, cosmological constant and dark energy
dominated universe and the specific value of $ \hat{w}$ depends on
the energy-matter content of the other three dimensions.

A codimension-2 braneworld model in six dimensions will be
discussed next. Results similar to the Gauss-Bonnet case
\cite{Bostock:2003cv}, {\it i.e.} four dimensional gravity for an
arbitrary energy momentum tensor, can be obtained if we include in
the action an induced gravity correction term instead. Again, in
the purely conical case, there appear to be no corrections to the
Einstein equations coming from the extra dimensions. The most
important observation is that the brane and bulk energy momentum
tensors are strongly related and any cosmological evolution on the
brane is dictated by the bulk
content~\cite{Papantonopoulos:2005ma}. We also see how this
correlation is relaxed in the case where bulk Gauss-Bonnet terms
and brane induced gravity terms are combined. Thus, the necessary
presence of extra curvature terms in the gravitational action in
order to give non-trivial gravitational dynamics on a
codimension-2 brane, leads to a realistic cosmological evolution
on the
 brane in the thin brane limit, only if a Gauss-Bonnet term is
included. However, let us note that the most physical way to
investigate
  the dynamics of codimension-2 branes, is by giving thickness to the brane
    \cite{Vinet:2004bk,Navarro:2004di}.

We will discuss in details the cosmological evolution of a conical
brane with both an induced gravity and a Gauss-Bonnet term added
in the higher dimensional gravity action
\cite{Papantonopoulos:2005nw}. For simplicity we will assume that
the only matter in the bulk is cosmological constant $\Lambda_B$.
We will solve the equations of motion evaluated on the brane and
assume that the integration of them in the bulk does not give rise
to pathologies ($i.e.$ singularities).

Firstly, we will discuss the isotropic cosmology, in which the
brane matter has to obey a tuning relation. The physical meaning
of this relation is that for any matter we put on the brane its
"image" should be present in the bulk. We will see that the
evolution of the system for $\Lambda_B=0$ tends to a fixed point
with $w=1/3$ and for $\Lambda_B>0$ to a fixed point with $w=-1$.
For $\Lambda_B<0$ the system has a runaway behaviour to $w \to
\infty$.

We will then relax the isotropy requirement for the metric
(keeping, however, the matter distribution isotropic) in order to
find whether the above matter tuning is an attractor or not. The
matter on the brane need not now satisfy the previous tuning
relation and the allowed regions of  initial values of the energy
density and pressure are significantly larger.  The analysis of
the dynamics of the system shows that line of isotropic tuning is
an attractor for $\Lambda_B \geq 0$ and thus the system
isotropises towards it. However, for values of $\Lambda_B$ which
give a realistic cosmological evolution of the equation of state,
the attractor property of the previous line is very weak and fine
tuning of the initial conditions is necessary in order to have an
evolution with acceptably small anisotropy. For $\Lambda_B < 0$
the system shows, as in the isotropic case, a runaway behaviour.

Finally we will discuss in brief, six-dimensional cosmological
models coming from supergravity theories. These models are
interesting because, they give conventional four-dimensional
cosmology with less arbitrariness in cosmological parameters and
less fine-tuning.
\def\thesection{\arabic{section}}
\section{A $(4+1)$ Cosmological Braneworld Model in
Six-Dimensional Spacetime}\label{intro6d}

\def\thesection{\arabic{section}.}

In this section we describe a braneworld cosmological model of
codimension-1 in six dimensions. This model is a generalization of
the existing braneworld models in five-dimensions. We study the
model following two different approaches. First we use normal
gaussian coordinates to describe a static 4-brane at fixed
position in a six-dimensional spacetime bulk. The six-dimensional
Einstein equations are derived and with the use of the appropriate
junction conditions the generalized Friedmann equation is given.
The cosmological evolution described by this generalized Friedmann
equation involves the usual three-dimensional scale factor $a(t)$
and the scale factor $b(t)$ describing the cosmological evolution
of the extra fourth dimension.

We consider next a dynamical brane moving in a bulk described by
six-dimensional static ``coordinates". In this case the dynamical
brane is moving on a geodesic which is given by the junctions
conditions. We derive the equations of motion of the brane which
for a brane observer describe the cosmological evolution on the
brane. We discuss the connection between the static and dynamical
brane models and the physical information that can be extracted
from these approaches.

\def\thesubsection{\arabic{section}-\arabic{subsection}}

\subsection{Static Brane in a Dynamical Bulk} \label{6ddynamic}

\def\thesubsection{\arabic{section}-\arabic{subsection}.}

 We look for a solution to the Einstein equations in six-dimensional spacetime with a metric of the form
\bea \label{metric6} ds^2 &=& -n^2(t,y,z) dt^2 + a^2(t,y,z)
d\Sigma_{k} ^2\nn \\ &+& b^2(t,y,z)  dy^2 + d^2(t,y,z) dz^2 \, ,
\eea where $d\Sigma_{k} ^2$ represents the 3-dimensional spatial
sections metric with $k=-1,\,0,\,1$ corresponding to the
hyperbolic, flat and elliptic spaces, respectively.

If the brane is fixed at the position $z_{0}$, then the total
energy-momentum tensor can be decomposed in two parts
corresponding to the bulk and the brane as
\begin{equation}\label{emtensor}
\tilde T^M _N = \breve T^{M(B)} _N + T^{M (b)} _N \, ,
\end{equation}
where the energy-momentum tensor on the brane is
\begin{equation}\label{brane}
T^{M (b)} _N = { {\delta (z-z_0)} \over {d}} \,diag \,(-\rho,
p,p,p,\hat p, 0) \, ,
\end{equation} where $\hat p$ is the pressure in the extra brane
dimension. We assume that there is no matter in the bulk and the
energy-momentum tensor of the bulk is proportional to the
six-dimensional cosmological constant.

The presence of the brane in $z_{0}$ imposes boundary conditions
on the metric: it must be continuous through the brane, while its
derivatives with respect to $z$ can be discontinuous at the brane
position. This means that the generated Dirac delta function in
the metric second derivatives with respect to $z$ must be matched
with the energy-momentum tensor components (\ref{brane}) to
satisfy the Einstein equations. The  Darmois-Israel conditions
are~\cite{Cuadros-Melgar:2005ex},
\begin{eqnarray}\label{israel}
{{[\partial_z a]} \over {a_0 d_0}} &=& -{{\kappa_{(6)} ^2} \over
4}
(p-\hat p + \rho) \, ,\nonumber \\
{{[\partial_z b]} \over {b_0 d_0}} &=& -{{\kappa_{(6)} ^2} \over
4}
\left\{ \rho - 3(p-\hat p) \right\} \, ,\\
{{[\partial_z n]} \over {d_0 n_0}} &=& {{\kappa_{(6)} ^2} \over 4}
\left\{ \hat p + 3 (p+ \rho) \right\} \, , \nonumber
\end{eqnarray}
where the subscript $(0)$ indicates quantities on the brane. The
energy conservation equation on the brane can be derived taking
the jump of the $(06)$ component of the Einstein equations and
using the junction conditions (\ref{israel}) and the corresponding
time derivatives.  We obtain
\begin{equation}\label{conserva6}
\dot\rho + 3 (p+\rho) {{\dot a_0}\over a_0} + (\hat p + \rho)
{{\dot b_0}\over b_0} =0 \, .
\end{equation}

To find the Friedmann equation we take the jump of the $(66)$
component of the Einstein equations and we use the fact that
\begin{equation}\label{prodsal}
[\partial_z f \, \partial_z g] = \#\partial_z f\# \, \,
[\partial_z g] + [\partial_z f] \,\, \#\partial_z g\# \, ,
\end{equation}
where $$\#f(y)\# = {{f(0^+) + f(0^-)}\over 2}\, ,$$ is the mean
value of the function $f$ through $y=0$, and we arrive to the
following equation
\begin{equation}\label{valmed}
{{\# \partial_z a \#}\over a_0} p = {1 \over 3} \rho {{\#
\partial_z n \#}\over n_0} - {1 \over 3} \hat p {{\# \partial_z b
\#}\over b_0} \, .
\end{equation}

Taking the mean value of the $(66)$ component of the Einstein
equations and using the junction equations, (\ref{valmed}) (we
have assumed a $Z_2$ symmetry), we arrive to the generalized
Friedmann equation in six-dimensions
\begin{eqnarray}\label{friedmanndist}
&~&\left( {{\ddot a_0}\over a_0} + {1\over 3} {{\ddot b_0}\over
b_0} + {{\dot a_0 \dot b_0}\over{a_0 b_0}} + {{\dot a_0 ^2}\over
{a_0
^2}} \right) = \nn \\
 &-& {{\kappa_{(6)} ^4}\over {32}} \left\{ \rho (\rho +2p +
{2\over 3} \hat p) + (p-\hat p)^2 \right\} -\nonumber \\&&-
2{k\over {a_0 ^2}}
  -{{\kappa_{(6)} ^2}\over{3 d_0 ^2} } \breve
T_{66}\, ,
\end{eqnarray} where we have assume $(\partial_{y}a)_{0}=0$, $(\partial_{y}^{2}a)_{0}=0$
and we have chosen $n_0=1$.

In the case of a 3-brane in a five-dimensional bulk, the first
integral of the space-space component of the Einstein equations,
with the help of the other equations can be done
analytically~\cite{5d} and this results in the Friedmann equation
on the brane with the dark radiation term as an integration
constant~\cite{maartens}. In our case the Einstein equations
 cannot be integrated analytically and
therefore, the usual form of the Friedmann equation on the brane
cannot be extracted from (\ref{friedmanndist}). Nevertheless, if
$a(t)$ and $b(t)$ are related, this equation will give the
cosmological evolution of the scale factor $a(t)$.

If $a(t)=b(t)=\cal{R}$(t) then (\ref{friedmanndist}) becomes
\begin{equation} \label{rsgen}
2 {{\ddot {\cal R}}\over {\cal R}} + 3 \left( {{\dot {\cal
R}}\over
    {\cal R}} \right)^2 = -3 {{\kappa_{(6)} ^4}\over {64}} \rho ^2 -
    {{\kappa_{(6)} ^4}\over {8}} \rho p - 3 {{k}\over {{\cal R}^2}} - {{\kappa_{(6)} ^2}\over {2 d_0 ^2}} \breve T^6 _6 \, .
    \end{equation}
This equation is the generalization of the Randall-Sundrum
Friedmann-like equation in six dimensions, and it has, as
expected, the $\rho^{2}$ term with a coefficient adjusted to six
dimensions. Note that this equation can easily be generalized to D
dimensions.

\def\thesubsection{\arabic{section}-\arabic{subsection}}

\subsection{Dynamical Brane in a Static Bulk}\label{6dstatic}

\def\thesubsection{\arabic{section}-\arabic{subsection}.}

We consider a 4-brane moving in a six-dimensional
Schwarzschild-AdS spacetime. The metric in the ``Schwarzschild"
coordinates can be written as
\begin{equation}\label{bhmetric}
ds^2 = - h(z) dt^2 + {z^2 \over l^2} d\Sigma_k ^2 + h^{-1} (z)
dz^2 \, ,
\end{equation}
where
\begin{equation}\label{spacediff}
d\Sigma_k ^2 = {{dr^2} \over {1-kr^2}} + r^2 d\Omega_{(2)} ^2 +
(1-kr^2) dy^2 \, ,
\end{equation}
and
\begin{equation}
h(z) = k + {z^2 \over l^2} - {M \over z^3} \,  .\label{hsch}
\\\end{equation}
Comparing the metric (\ref{metric6}) with the metric
(\ref{bhmetric}) we can make the following identifications
\begin{eqnarray}\label{identifica}
n(z) &=& \sqrt{h(z)} \, , \nonumber \\
a(z)&=&b(z)=z/l~, \nn \\ d(z)&=& \sqrt{h^{-1}(z)} \, .
\end{eqnarray}
The two approaches of a static brane in a dynamical bulk described
in Sec.~\ref{6ddynamic} and of a moving 4-brane in a static bulk
are equivalent~\cite{Mukohyama:1999wi}. To prove this equivalence,
 consider the Darmois-Israel conditions for a moving brane,
$[K^\mu _\nu] = -\kappa_{(6)} ^2 \left(T^\mu _\nu - {1 \over 4} T
h^\mu _\nu \right)$, with $h^\mu _\nu$ being the induced metric on
the brane; analogously to the case in 5 dimensions one can obtain
\begin{eqnarray}
{{-h'-2 \ddot{\cal R}}\over {\sqrt{h+ \dot{\cal R}^2}}} &=&
-\kappa_{(6)}
^2 ({3\over 4} \rho -p) \, , \label{equ1} \\
-2 {{\sqrt{h+\dot{\cal R}^2}} \over {\cal R}} &=&
{{\kappa_{(6)}^2}
  \over 4} \rho \, ,\label{equ2}
\end{eqnarray} where we have defined $z=\mathcal{R}(t)$.
Combining these two equations and using (\ref{hsch}) we find that
\begin{equation} \label{brfriedeq}
2 {{\ddot {\cal R}}\over {\cal R}} + 3 \left( {{\dot {\cal
R}}\over
    {\cal R}} \right)^2 = -3 {{\kappa_{(6)} ^4}\over {64}} \rho ^2 -
    {{\kappa_{(6)} ^4}\over {8}} \rho p - 3 {{k}\over {{\cal R}^2}} -
    {5\over {l^2}} \, .
\end{equation}
Comparing with (\ref{rsgen}) we see that
\begin{eqnarray}
{\kappa_{(6)} ^2} \breve T^6 _6 ={{\kappa_{(6)} ^2}\over {d ^2}}
\breve T_{66} = {{10}\over l^2} \nonumber \\
\Rightarrow l^{-2} = -{{\kappa_{(6)} ^2}\over {10}} \Lambda_{_{6}}
\, ,
\end{eqnarray}
with $\breve T^6 _6 =-\Lambda_{6}$ the bulk cosmological constant,
and $l$ the size of the AdS space. Therefore, a brane observer
describes the cosmological evolution of a four-dimensional
universe with the Friedmann equation (\ref{brfriedeq}) with
$\mathcal{R}$ parameterizing the motion of the brane in the $z$
direction.

To describe the motion of the 4-brane with $a(z)\neq b(z)$ we can
write the metric (\ref{bhmetric}) as
 \begin{equation}\label{bhmetric1}
ds^2 = - n(z)^{2} dt^2 +a^{2}(z)d\Sigma_3 ^2 +b(z)^{2}dy^2+
d^{2}(z)dz^2\, .
\end{equation} We
denote the position of the brane
  at any bulk time $t$ by $z = {\cal R}(t)$ as before. Then, an observer on the brane defines the proper time
from the relation \be n^2(t,{\cal R}(t))\dot{t}^{2}-d^2(t,{\cal
R}(t)) \dot{\cal R}^2=1~, \label{dtdtau} \ee which ensures that
the induced metric on the brane will be in FRW form
\begin{eqnarray}\label{induced}
ds^2_{induced} &=& - \left[ n^2(t,{\cal R}(t)) \dot
t^2-d^2(t,{\cal R}(t)) \dot
{\cal R}^2 \right] d\tau ^2 \nonumber \\ &&+ \,\,a^2(t,{\cal R}(t)) d\Sigma^2 _{(3)}+b^{2}(t,{\cal R}(t))dy^2 \nonumber \\
&=& -d\tau^2 + a^2(t,{\cal R}(t)) d\Sigma^2 _{(3)}\nn
\\&+& b^{2}(t,{\cal R}(t))dy^2 \, ,
\end{eqnarray}
where the dot indicates derivative with respect to the brane time
$\tau$.

Introducing an energy-momentum tensor on the brane
\begin{equation}
\hat{T}_{\mu\nu}=h_{\nu\alpha}T^{\alpha}_{\mu}-\frac{1}{4}T
h_{\mu\nu}, \end{equation} where
$T^{\alpha}_{\mu}=diag(-\rho,p,p,p,\hat{p})$, the Darmois-Israel
conditions become
\begin{equation}
[K_{\mu\nu}]=-\kappa^{2}_{(6)}\hat{T}_{\mu\nu}~,
\end{equation} where
$K_{\mu\nu}$ is the extrinsic curvature tensor. These give the
equations of motion of the brane
\begin{eqnarray}
\frac{d^{2} \dot{d} \dot{\cal R}^{3}-d\ddot{\cal R}}
{\sqrt{1+d^{2}\dot{\cal R}^{2}}}&-&\frac{\sqrt{1+d^{2}\dot{\cal
R}^{2}}}{n}\Big{(}d\dot{n}\dot{\cal R}+\frac{\partial_{z}n}{d}\nn
\\ &-&(d\partial_{z}n-n\partial_{z}d)\dot{\cal{R}}^{2}
\Big{)}\nonumber
\\ &=&
-\frac{\kappa^{2}_{(6)}}{8}\Big{(}3(\rho+p)+\hat{p} \Big{)}~,\label{dyneqs} \\
\frac{\partial_{z}a}{ad}\sqrt{1+d^{2}\dot{\cal R}^{2}}&=&
-\frac{\kappa^{2}_{(6)}}{8}\Big{(}\rho+p-\hat{p} \Big{)}~, \label{const1} \\
 \frac{\partial_{z}b}{bd}\sqrt{1+d^{2}\dot{\cal R}^{2}}&=&
-\frac{\kappa^{2}_{(6)}}{8}\Big{(}\rho-3(p-\hat{p})\label{const2}
\Big{)}~.
\end{eqnarray}
Notice that if $a=b={\cal R}/l$, we recover equations (\ref{equ1})
and (\ref{equ2}), corresponding to the Schwarzschild-AdS
spacetime.

Equation (\ref{dyneqs}) is the main dynamical equation describing
the movement of the brane-universe in the six-dimensional bulk,
while a combination of (\ref{const1}) and (\ref{const2}) acts as a
constraint relating $a$ and $b$ (remember that for a brane
observer $a$ and $b$ are static, depending only on z)\be
a=\mathcal{A}~b^{(\rho+p-\hat{p})/(\rho-3(p-\hat{p}))}\label{constrel},\ee
where $\mathcal{A}$ is an integration constant. This means that
 the relative
cosmological evolution of $a$, the scale factor of the
three-dimensional physical universe and $b$, the scale factor of
the extra dimension, depends on the dynamics of the energy-matter
content of the brane-universe.

Another interesting observation is that, a brane observer can
measure the departure from full six-dimensional spherical symmetry
of the bulk using the quantities $a$ and $b$. If $a(z)=b(z)=z/l$
then the symmetry of the bulk is $S^{4}$ having a six-dimensional
Schwarzschild-(A)dS black hole solution. If $a(z)\neq b(z)$ fixing
$a$ to be $a(z)=z/l$, because of (\ref{constrel}) $b(z)$ is given
by \be b=\left\{ \frac{1}{\mathcal{A}}
\Big{(}\frac{z}{l}\Big{)}\right\}^{(\rho-3(p-\hat{p}))/(\rho+p-\hat{p})}
\label{inva} \ee and we expect the topology of the bulk to be
$S^{3}\times {\cal M}$,  ${\cal M}$ being a compact or non-compact
manifold. There are no analytical solutions in the six-dimensional
spacetime with such topology and the reflection of this on the
brane is the difficulty of the brane equations
(\ref{dyneqs})-(\ref{const2}) to be integrated analytically. Note
also that because of (\ref{inva}), a change in the topology of the
bulk is triggered by the dynamics of the matter distribution on
the brane.

It is also interesting to further explore the relation we found
between the cosmological evolution of a higher-dimensional
brane-universe with the static properties of the bulk. The
cosmological evolution on a higher-dimensional brane-universe is
related to a topology change of the bulk during the evolution, and
this relation might lead to a better understanding of the
Gregory-Laflamme~\cite{Gregory:1993vy} instabilities of
higher-dimensional objects.

\def\thesection{\arabic{section}}

\section{The Cosmological Evolution of a (4+1)  Brane-Universe}

\def\thesection{\arabic{section}.}

To study the cosmological evolution of the four-dimensional
brane-universe, we made the following assumptions for the initial
conditions and the matter distribution on the brane. We assume
that the universe started as a four-dimensional one at the Planck
scale, all the dimensions were of the Planck length and the matter
was isotropically distributed. In this case the cosmological
evolution is described by the generalized Friedmann equation
(\ref{rsgen}). Then an anisotropy was developed in the sense that
$\hat{p}=Qp$ with $Q\neq 1$. The cosmological evolution is now
described by (\ref{friedmanndist}) supplemented with the
constrained equation (\ref{constrel}) and the matter distribution
on the brane is given
by the equations of state \bea p&=&w \rho~,\label{eqstq} \\
\hat{p}&=& \hat{w} \rho~. \label{eqstqhat} \eea Using
(\ref{constrel}) and relations (\ref{eqstq}) and (\ref{eqstqhat})
the generalized Friedmann equation (\ref{friedmanndist}) becomes
\bea
\frac{\hat{w}-w}{1+\hat{w}}~\dot{H}_{a}&+&\frac{(1+\hat{w})(-3w+2\hat{w}-1)+3(1+\hat{w})^{2}}{(1+\hat{w})^{2}}
~H^{2}_{a} \nn \\
&+&\frac{2w-\hat{w}+1}{(1+\hat{w})^{2}}\frac{\dot{\rho}}{\rho}~H_{a}+\frac{2+\hat{w}}{3(1+\hat{w})^{2}}
\frac{\dot{\rho}^{2}}{\rho^{2}}\nn
\\ &-&\frac{\ddot{\rho}}{3(1+\hat{w})\rho} \nn \\ &=&
 - {{\kappa_{(6)} ^2}\over {32}} \left\{ 1+2w+\frac{2}{3}\hat{w}+(w-\hat{w})^{2} \right\}
 \rho^{2}\nn \\ &-& 2{k\over {a^2}}
  +{{\kappa_{(6)} ^2}\over{3 } }
\Lambda_{6}\, . \eea Using the conservation equation
(\ref{conserva6}) to eliminate $\rho$ and its derivatives from the
above equation, the cosmological evolution of the
three-dimensional scale factor $a(t)$ is given by the equation
\bea \Big{[}1+\frac{B}{3}\Big{]}\ddot{a}a^{2C+1}&+&\Big{[}
\frac{B^{2}}{3}+\frac{2B}{3}+1 \Big{]}\dot{a}^{2}a^{2C}\nn \\&+&
\frac{\kappa_{(6)} ^2} {32} a^{2} \Big{[}
 1+2w+\frac{2}{3}\hat{w}+(w-\hat{w})^{2} \Big{]} \nn \\
 &-& a^{2C} \Big{[} a^{2} \frac{\kappa_{(6)}^2}{3} \Lambda_{6}-2k
 \Big{]}=0~, \label{aevolut} \eea  where the constants $B$ and $C$ are given by
  \bea B&=&\frac{1-3w+3\hat{w}}{1+w-\hat{w}}~, \\
C&=&3(1+w)+B(\hat{w}+1)~, \eea while the $b(t)$ scale factor is
\be
 b(t)=a(t)^{B}~. \label{bevolut}
\ee We  made a numerical analysis of equations (\ref{aevolut}) and
(\ref{bevolut}) and studied the time evolution of the two scale
factors for different backgrounds and spatial brane-curvature. We
 allowed for all possible forms of energy-matter on the physical
three dimensions ($w$=0, 1/3, -1/3) and also for the possibility
of dark energy ($w$=-1), leaving $\hat{w}$ as a free parameter.

For the scale factor $b(t)$ to be small compared to the scale
factor $a(t)$, the constant $B$ in (\ref{bevolut}) should be
negative. In Table 1 we give the allowed range of values of
$\hat{w}$ for various values of $w$. These values in turn were
used to plot the time evolution of the scale factors $a(t)$ and
$b(t)$ using (\ref{aevolut}) and (\ref{bevolut}) respectively. The
criterion for the acceptance of a solution is to give a growing
evolution of $a(t)$ and a decaying and freezing out evolution for
$b(t)$. The results for various choices of the parameters of the
model are presented in the following table

\begin{table}[h]
\begin{center}
\begin{tabular}{|l|l|r|}
 \hline
$~~~w$ & $~~~~~~~~~~\hat{w}$ \\
\hline
$\,\,-1$  & $>0\,\,\ ~~~or \,\,\,\ <-4/3 $ \\
\hline
$\,\,\,\,~0$ & $>1\,\,\ ~~~or \,\,\ <-1/3 $\\
\hline
$\,\,\,1/3$ & $>4/3\,\,\ or \,\,\,\ <0 $\\
\hline
$-1/3$ & $>2/3\,\,\ or \,\,\,\ <-2/3 $\\
\hline
\end{tabular}
\end{center}
\caption{The allowed values of $\hat{w}$ for $B$ to be negative.}
\end{table}

The numerical analysis showed~\cite{Cuadros-Melgar:2005ex} that
in all the cases considered, $\hat{w}$ is negative in the range of
values given in Table 1 for all acceptable solutions, indicating
the need of dark energy to suppress the extra fourth dimension
compared to the three other dimensions.

\def\thesection{\arabic{section}}

 \section{Codimension-2 Braneworld Model with Induced Gravity in Six Dimensions}

 \def\thesection{\arabic{section}.}

We will next discuss a codimension-2 braneworld model in six
dimensions. We will first consider a six-dimensional theory with
general bulk dynamics encoded in a Lagrangian ${\cal L}_{Bulk}$
and a 3-brane at some point $r=0$ of the two-dimensional internal
space with general dynamics ${\cal L}_{brane}$
 in its world-volume. If we include an induced curvature term localized at the position of the brane,
  the total action is written as:
\bea
 {\cal S}&=&{M^4_6 \over 2}\left[\int d^6x \sqrt{G}R^{(6)}+r_c^2\int d^4x
\sqrt{g}R^{(4)} ~{\d(r) \over 2 \pi L}\right]\nn \\  &+& \int d^6x
{\cal L}_{Bulk} + \int d^4x {\cal L}_{brane}~{\d(r) \over 2 \pi
L}~.\label{inaction} \eea In the above action, $M_6$ is the
six-dimensional Planck mass,
 $M_4$ is the four-dimensional one and $r_c=M_4/M_6^2$ the cross over scale between
  four-dimensional and six-dimensional gravity.
 The above induced term has been
written in the particular coordinate system in which the metric is
\be ds_6^2=g_{\m\n}(x,r)dx^\m dx^\n+dr^2+L^2(x,r)d\th^2
\label{metric}~, \ee where $g_{\mu\nu}(x,0)$ is the braneworld
metric and $x^{\mu}$ denote four non-compact dimensions,
$\mu=0,...,3$, whereas $r,\th$ denote the radial and angular
coordinates of the two extra dimensions (the r direction may
 or may not  be compact and the $\th$ coordinate ranges form $0$ to $2\pi$).
Capital $M$,$N$ indices will take values in the six-dimensional
space. Note, that we have assumed that there exists an azimuthal
symmetry in the system, so that both the induced four-dimensional
metric and the function $L$ do not depend on $\th$.
 The normalization of the $\d$-function is the one discussed in \cite{Leblond:2001xr}.

To obtain the braneworld equations we expand the metric around the
brane as \be L(x,r)=\beta(x)r+O(r^{2})~. \ee At the boundary of
the internal two-dimensional space where the 3-brane is situated,
the function $L$ behaves as $L^{\prime}(x,0)=\beta(x)$, where a
prime denotes derivative with respect to $r$. As we will see in
the following, the demand that the space in the vicinity of the
conical singularity is regular, imposes the supplementary
conditions that $\de_\m \b=0$ and $\partial_{r}g_{\mu\nu}(x,0)=0$.

The Einstein equations which are derived from the above action in
the presence of the 3-brane are \bea &~&
 G^{(6)N}_M + r_c^2
G^{(4)\n}_\m \d_M^\m \d^N_\n {\d(r) \over 2 \pi L}\nn \\ &=& {1
\over M_6^4} \left[T^{(B)N}_M + T^{(br)\n}_\m \d_M^\m \d^N_\n
{\d(r) \over 2 \pi L}\right]~, \label{einsequat} \eea with
$G^{(6)N}_M$ and $G^{(4)\n}_\m$ the six-dimensional and the
four-dimensional Einstein tensors respectively,   $T^{(B)N}_M$ the
bulk energy momentum tensor and $T^{(br)\n}_\m$ the brane one.

We will now use the fact that the second derivatives of the metric
functions contain $\d$-function singularities at the position of
the brane. The nature of the singularity then gives the following
relations \cite{Bostock:2003cv} \bea
{L'' \over L}&=&-(1-L'){\d(r) \over L}+ {\rm non-singular~terms}~,\\
{K'_{\m\n} \over L}&=&K_{\m\n}{\d(r) \over L}+ {\rm
non-singular~terms}~. \eea From the above singularity expressions
we can  match the singular parts of the Einstein equations
(\ref{einsequat}) and get the following ``boundary" Einstein
equations \be G^{(4)\n}_\m|_0 = {1 \over r_c^2 M_6^4}T^{(br)\n}_\m
+{2\pi \over r_c^2} (1-\b)\d_\m^\n +{2 \pi L \over
2r_c^2}(K_{\m}^\n-\d_\m^\n K)|_0 \label{indeq}~, \ee where
$K_{\m\n}$ is the extrinsic curvature and we denote by $|_0$ the
value of the corresponding function at $r=0$.

We will now make the assumption that the singularity is purely
conical. In the opposite case, there would be curvature
singularities $R^{(6)} \propto 1/r$, because in the Ricci tensor
$R^{(6)}_{\mu\nu}$ there are terms of the form
\cite{Bostock:2003cv} \be
R^{(6)}_{\mu\nu}=-\frac{1}{2}\frac{L^{\prime}}{L}\partial_{r}g_{\mu\nu}+...=-\frac{\partial_{r}g_{\mu\nu}}{2r}
+\mathcal{O}(1)~, \ee
 which in the vicinity of $r=0$ are singular
if $\partial_{r}g_{\mu\nu}(x,0)\neq0$. The absence of this type of
singularities imposes the requirement that $K_{\m\n}|_0=0$. Then,
the Einstein equations (\ref{indeq}) reduce to \be
G^{(4)}_{\m\n}|_0= {1 \over r_c^2 M_6^4}T^{(br)}_{\m\n} +{2 \pi
\over r_c^2} (1-\b)g_{\m\n}|_0 \label{conindeq}~. \ee

The four-dimensional Einstein equations (\ref{conindeq}) describe
the gravitational dynamics on the
brane~\cite{Papantonopoulos:2005ma}. The effective
four-dimensional Planck mass and cosmological constant are simply
\bea
 M^{2}_{Pl}~=~M^2_4&=& r_c^2 M_6^4~,\\
\Lambda_{4}&=&\lambda-2\pi M_6^4 (1-\b)~, \eea where $\lambda$ is
the contribution of the vacuum energy of the brane fields. The
normalization of $\Lambda_4$ is defined by the convention that the
four dimensional Einstein equation reads \mbox{$G_{\m\n}={1 \over
M_{Pl}^2}(T_{\m\n}-\Lambda_4 g_{\m\n})$}. Note that in contrast to
the case of \cite{Bostock:2003cv}, the four dimensional Planck
mass is independent of the deficit angle.

 Furthermore, it is interesting to see  that contrary to
the five-dimensional case, the induced gravity term in six
dimensions does not introduce any correction terms, apart from a
cosmological term,  in the four-dimensional Einstein equations on
the brane, unless singularities of other type than conical  are
allowed in the theory, and a regularization scheme is employed. In
the latter case, the last term of the right hand side of
(\ref{indeq}) would provide information of the bulk physics. This
absence of corrections in the purely conical case,  is exactly
what happens also in the case of the bulk Gauss-Bonnet theory
\cite{Bostock:2003cv}.

What is important to note at this point, is that although we have
found a ``boundary'' Einstein equation, there is more information
about the dynamics of the theory  contained in the full
six-dimensional Einstein equations.
In~\cite{Papantonopoulos:2005ma} it was showed that the energy
momentum tensor of the bulk is strongly related to the energy
momentum tensor of the brane via the relation
 \be
T^{(B)r}_r|_0={1 \over 2r_c^2}\left[T^{(br)\m}_\m+8 \pi M_6^4
(1-\b)\right]~.\label{bulkbranem} \ee This equation constitutes a
very strong tuning relation between brane ($T^{(br)\m}_\m$) and
bulk ($T^{(B)r}_r|_0$) matter. It shows that, in order to have
some cosmological evolution on the brane (\ie time dependent
$T^{(br)}_{\m\n}$) and since $\b$ is constant,  the bulk content
should evolve as well in a precisely tuned way.

We can compare the above result with what is happening in five
dimensions. The action of a general five-dimensional theory with a
3-brane at the point $r=0$ of the extra dimension, and with an
induced curvature term localized on it, is \bea
 {\cal S}&=&{M^3_5 \over 2}\left[\int d^5x \sqrt{G}R^{(5)}+r_c\int d^4x
\sqrt{g}R^{(4)} ~\d(r)\right]\nn \\  &+& \int d^5x {\cal L}_{Bulk}
+ \int d^4x {\cal L}_{brane}~\d(r) \label{action}~. \eea
 In the
above action, $M_5$ is the five-dimensional Planck mass, $M_4$ is
the four dimensional one and $r_c=M_4^2/M_5^3$ the cross over
scale of the five-dimensional theory. The above induced term has
been written in the particular coordinate system in which the
metric is written as \be ds_5^2=g_{\m\n}(x,r)dx^\m dx^\n+dr^2~,
\ee where the $x^{\mu}$ denote the usual four non-compact
dimensions, $\mu=0,...,3$, whereas $r$ denotes the radial extra
coordinate and capital $M$,$N$ indices will now take values in the
five-dimensional space.

The Einstein equations which are derived from the action
(\ref{action}) in the presence of the 3-brane are \bea
 &~&G^{(5)N}_M +
r_c G^{(4)\n}_\m \d_M^\m \d^N_\n \d(r)=\nn \\ &~&{1 \over M_5^3}
\left[T^{(B)N}_M+T^{(br)\n}_\m \d_M^\m \d^N_\n
\d(r)\right]~,\label{5deineq} \eea
 with $T^{(B)N}_M$ the bulk
energy momentum tensor and $T^{(br)\n}_\m$ the brane one.
Performing a similar analysis like the six-dimensional
case~\cite{Papantonopoulos:2005ma} we get \be
 R^{(4)}|_0+{1
\over 4}(K_\m^\n K^\m_\n-K^2)|_0=- {2 \over M_5^3}T^{(B)r}_r|_0~.
\ee

From the above equation we see that the bulk matter content does
not necessarily dictate the brane cosmological evolution. This is
because the extrinsic curvature on the brane $K_{\m \n}$ can be
non-trivial and it is this one which plays the most crucial role
in the cosmology. In other words, in five dimensions it is the
freedom of the brane to bend in the extra dimension which makes
the evolution not tuned to the bulk matter content. The absence of
such bending in six dimensions (imposed by singularity arguments)
gives the bulk the crucial role for how the brane evolves.

We can also introduce a Gauss-Bonnet term in the six-dimensional
action (\ref{inaction}) and see how the above results are
modified. In this case the action (\ref{inaction}) is augmented by
the term \be S_{GB}={M_6^4 \a \over 2}\int d^6x
(R^{(6)~2}-4R^{(6)~2}_{MN}+R^{(6)~2}_{MNK\Lambda})~. \ee
 Then the
variation of the above action introduces an extra term in the left
hand side of the Einstein equations (\ref{einsequat}), \bea
 H_M^N&=&-\a \Big{[}{1
\over 2}\d_M^N (R^{(6)~2} -4R^{(6)~2}_{K\Lambda}+R^{(6)~2}_{ABK
\Lambda})\nn \\ &-& 2R^{(6)}R^{(6)N}_{M}+
4R^{(6)}_{MP}R^{NP}_{(6)}\phantom{{1 \over 2}}\nn \\ &+&
4R^{(6)~~~N}_{KMP}R_{(6)}^{KP}  - 2R^{(6)}_{MK\Lambda
P}R_{(6)}^{NK\Lambda P}\Big{]} \label{gbterm}. \eea Equating the
singular terms of the Einstein equations by the standard procedure
of section 2, and demanding that the singularity is purely
conical, we obtain the following
 ``boundary" Einstein equations
 \bea
G^{(4)}_{\m\n}|_0&=&{1 \over M_6^4 (r_c^2+8\pi
(1-\b)\a)}T^{(br)}_{\m\n}\nn \\ &+& {2\pi (1-\b) \over r_c^2+8\pi
(1-\b)\a}g_{\m\n}|_0 \label{einsteincomb}~. \eea
 Equation
(\ref{einsteincomb}) describes the gravitational dynamics on a
codimension-2 brane when both induced gravity and Gauss-Bonnet
correction terms are present. The effective four-dimensional
Planck mass and cosmological constant are simply \bea
 M^{2}_{Pl}&=&M_6^4 (r_c^2+8\pi (1-\b)\a)~,\\
\Lambda_{4}&=&\lambda-2\pi M_6^4 (1-\b)~, \eea where $\lambda$ is
the brane tension.  Note that the Planck mass this time can depend
on the deficit angle. This is an effect of solely the bulk
Gauss-Bonnet term.

Evaluating the the $(rr)$ component of the Einstein equation at
the position of the brane $r=0$ we obtain the following
relation~\cite{Papantonopoulos:2005ma}
 \be
 R^{(4)}|_0+\a
(R^{(4)~2}-4R^{(4)~2}_{\k\l}+R^{(4)~2}_{\a\b\k\l})|_0 =-{2 \over
M_6^2}T^{(B)r}_r|_0 \label{rrcomb}~. \ee

From (\ref{rrcomb}) we see that there can be no relation between
the extra dimensional component $T^{(B)r}_r|_0$ of the bulk energy
momentum tensor  at the position of the brane with the brane
energy momentum tensor $T^{(br)\m}_\m$. This is due to the
appearance of the Riemann curvature which cannot be evaluated from
previous equations (the Ricci tensor and scalar can be substituted
from (\ref{einsteincomb})). Instead, using  (\ref{einsteincomb})
and (\ref{rrcomb}), one can {\it solve} for
$R^{(4)~2}_{\a\b\k\l}|_0$  as a function of the brane and bulk
matter at the position of the brane.

\def\thesection{\arabic{section}}

\section{Cosmological Evolution of a Conical Codimension-2 Braneworld Model}

\def\thesection{\arabic{section}.}

We consider a six-dimensional theory with general bulk dynamics
encoded in a Lagrangian ${\cal L}_{Bulk}$ and a 3-brane at some
point $r=0$ of the two-dimensional internal space with general
dynamics ${\cal L}_{brane}$
 in its world-volume. The gravitational dynamics is described as we discuss in the previous sections
  by a Gauss-Bonnet term
 in the bulk and an induced four-dimensional curvature term localized at the position of the
 brane. Then the total action is written as
\bea
 {\cal S} &=& {M^4_6 \over 2}\Big{\{} \int d^6x
 \sqrt{-g^{(6)}}[R^{(6)}+\a(R^{(6)~2} \nn \\
 &-& 4R^{(6)~2}_{MN}+R^{(6)~2}_{MNK\Lambda})]+
  r_c^2\int d^4x
\sqrt{-g}R^{(4)} ~{\d(r) \over 2 \pi L} \Big{\}}\nn \\ &+& \int
d^6x {\cal L}_{Bulk} + \int d^4x {\cal L}_{brane}~{\d(r) \over 2
\pi L}~.\label{inactionnew} \eea The full equations of motion that
are derived from the above action using the metric (\ref{metric})
are \bea &~& G^{(6)N}_M + r_c^2 G^{(4)\n}_\m \d^\m_M \d_\n^N
{\d(r) \over 2 \pi L}  -\a H_M^N\nn \\&=&{1 \over M_6^4}
\left[T^{(B)N}_M+T^{(br)\n}_\m \d_M^\m \d^N_\n {\d(r) \over 2 \pi
L}\right], \label{einsteineq}\eea
 with $G_M^{(6)~N}=R^{(6)~N}_M-{1 \over
2}R^{(6)} \d_M^N$ the six-dimensional Einstein tensor,
$G_\m^{(4)~\n}=R^{(4)~\n}_\m-{1 \over 2}R^{(4)} \d_\m^\n$ the
four-dimensional Einstein tensor and $H_M^N$ is given by
(\ref{gbterm}).

In order that there are no curvature singularities more severe
than conical,
 we will impose certain conditions on the value of the extrinsic curvature $K_{\m\n}=g'_{\m\n}$ on the
 brane, where the prime denotes derivative with respect to $r$,
  and on the expansion coefficients of the  function $L$
\be L=\b_1(x) r +\b_2(x) r^2 + \b_3(x) r^3 +\dots \ee
 These
conditions read \cite{Bostock:2003cv} \bea
K_{\m\n}|_{r=0}&=&0~,\label{condition1}\\
\b_1={\rm const.}~~~~&{\rm and}&~~~~\b_2=0~. \eea

Imposing these conditions and keeping only the finite part in
$L''/L$, the Einstein equations (\ref{einsteineq}) can be
evaluated at $r=0$. The effective Einstein equations on the brane
(obtained by equating the $\d$-function parts of the Einstein
equations) are  \be G^{(4)\n}_\m={1 \over M_{Pl}^2
}\left[T^{(br)\n}_\m -\Lambda_4\d_\m^\n \right]~,
\label{4deins}\ee with $M_{Pl}^2=M_6^4 [r_c^2+8\pi (1-\b_{1})\a]$
and $\Lambda_{4}=-2\pi M_6^4 (1-\b_{1})$. The various components
of the bulk Einstein equations evaluated at $r=0$ are given in the
following:

The $(\m\n)$ component \bea &~&G^{(4)\n}_\m -{1 \over 2}K_\m^{\n ~
'}+{1 \over 2}\d_\m^\n \Big{(}K'+2{L'' \over L}\Big{)}\nonumber
\\ &&-\a \Big{[} {1 \over 2}\d_\m^\n \Big{(}R^{2}-4
R_{\k\la}^{2}+R_{\k\la\xi\s}^{2}\nonumber \\ &&+4K^{\k~'}_\la
R_\k^\la-2 K' R  -4{L'' \over L}R \Big{)}  \nonumber \\
&&-2RR_\m^\n+4 R_{\m \k}R^{\n \k}+4R^{~~~~\n}_{\k \m
\lambda}R^{\k\lambda}\nonumber \\ &&-2R_{\m \k\lambda \rho}R^{\n
\k\lambda \rho} \nonumber \\ &&+4{L'' \over L}R^\n_\m
+2K'R^\n_\m+K_\m^{\n~'} R\nonumber \\ &&-2K_\k^{\n~'}
R^\k_\m-2g_{\m\la}K^{\la~'}_\k R^{\n\k}\Big{]}= {1 \over
M_6^4}T^{(B)\n}_\m. \label{mncomp} \eea

 The $(rr)$ component
\be
 -{1 \over 2}R -\a {1 \over 2}(R^{2}-4
R_{\m\n}^{2}+R_{\m\n\k\lambda}^{2})= {1 \over M_6^4} T^{(B)r}_r~.
\label{rrcomp} \ee

The $(\th \th) -(rr)$ component \be {1 \over 2}K' +\a \left( K'R -
2K_\n^{\m~'}R_\m^\n\right)={1 \over M_6^4} (T^{(B)\th}_\th -
T^{(B)r}_r)~. \label{ththcomp}\ee

The ($\m,r$) component \be T^{(B)r}_\m =0 ~.\ee

In the following, we will study the above equations in a time
dependent background, assuming that the bulk consists of a pure
cosmological constant $T^{(B)N}_M=- \La_B \d_M^N$ and that the
matter content of the brane is an isotropic fluid with
$T^{(br)\n}_\m=\d_\m^\n ~{\rm diag}(-\r_b,P_b,P_b,P_b)$.

\def\thesubsection{\arabic{section}-\arabic{subsection}}

\subsection{The Constrained Isotropic Case}

\def\thesubsection{\arabic{section}-\arabic{subsection}.}

We are interested in the cosmological evolution of a flat
isotropic brane-universe, therefore we will consider the following
time dependent form of the metric (\ref{metric}) \be ds^2=
-N^2(t,r)dt^2 +A^2(t,r)d\vec x^2 + dr^2 + L^2(x,r) d\th^2~.
\label{timmetric}\ee
 We can use the gauge freedom to fix
$N(t,0)=1$, while we define $A(t,0)\equiv a(t)$. The curvature
singularity avoidance condition (\ref{condition1}) we imposed,
dictates that $N'(t,0)=A'(t,0)=0$, while the second derivatives of
these metric functions are unconstrained.

For this ansatz, the Einstein equations (\ref{4deins}) and
(\ref{rrcomp}) give \bea 3{\dot{a}^2 \over a^2}&=&{\rho_{b}+\La_4
\over M_{Pl}^2}~,\label{1fried}
\\
2{\ddot{a} \over a } + {\dot{a}^2 \over a^2}  &=&{-P_{b}+\La_4
\over M_{Pl}^2 }~,\label{2fried} \\
 3  \left( {\ddot{a} \over a}+{\dot{a}^2 \over a^2} \right) &+&12
\a {\ddot{a}\dot{a}^2 \over a^3}= {\La_B \over
M_6^4}~.\label{friedconstraint} \eea

The equations (\ref{1fried}) and (\ref{2fried}) are the usual
Friedmann and Raychaudhuri equations of a four-dimensional
universe with a scale factor $a$, while the third equation
(\ref{friedconstraint}) appears because of the presence of the
bulk and acts as a constraint between the matter density and
pressure on the brane. To see this, a simple manipulation of the
above equations gives
 \bea
-{\La_B \over M_6^4}&=&\left({1 \over 2}+{2 \over 3}\a{\La_4 \over
M_{Pl}^2}\right) {3P_b-\r_b \over M_{Pl}^2}\nn \\ &-&2{\La_4 \over
M_{Pl}^2}\left( 1+ {2 \over 3}\a{\La_4 \over M_{Pl}^2} \right)\nn
\\ &+& {2 \over 3}\a {\r_b (3P_b+\r_b ) \over M_{Pl}^2}
\label{Priso}~, \eea which shows a precise relation between
$\La_B$, $\r_b$ and $P_b$. To simplify the equations, we assume
that the vacuum energy (tension) of the brane cancels the
contribution $\La_4$ induced by the deficit angle, \ie
 \be \r_b=-\La_4 +
\r_m~~~{\rm and} ~~~ P_b=\La_4  + P_m~,\label{relations} \ee
 with $P_m=w_c\r_m$. Then
the above equations read
 \bea
&&3{\dot{a}^2 \over a^2}={\rho_m \over M_{Pl}^2}~, \label{1rstfr}\\
&&2{\ddot{a} \over a } + {\dot{a}^2 \over a^2}  =-w_c{\r_m \over
M_{Pl}^2 }~,\label{2ndfr} \eea while the constraint equation
becomes
 \be
-{\La_B \over M_6^4}={\r_m \over M_{Pl}^2}\left[{1 \over
2}(3w_c-1)+{2 \over 3}(3w_c+1)\a{\r_m \over M_{Pl}^2}
   \right]~.
\label{isocon} \ee

Using the metric (\ref{timmetric}), the Einstein equations
(\ref{mncomp}) and (\ref{ththcomp})  can be solved for the second
$r$-derivatives of the metric, as functions of the matter content
on the brane, and in principle they can give us information about
the structure of the bulk at $r=0$ \bea
 {A'' \over A}&=&{ {1 \over 4}\left(1+4\a
{\La_B \over M_{6}^4}\right)(w_c+1){\r_m \over M_{Pl}^2}  \over
1- 2\a
{\r_m \over M_{Pl}^2}\left[w_c-1+2\a {\r_m \over M_{Pl}^2}(w_c+1)(3w_c-1)   \right]   } \label{A''sol}\nn \\
{N'' \over N}&=&3 {A'' \over A}~ {4\a w_c{\r_m \over M_{Pl}^2} -1 \over 4 \a {\r_m \over M_{Pl}^2} +1 }~,
 \label{N''sol}\nn \\
{L'' \over L}&=&{1 \over 1+ 4\a {\r_m \over M_{Pl}^2}}\Big{[}
{\r_m \over M_{Pl}^2} -{\La_B\over M_{6}^4}\nn \\  &-3& {A'' \over
A}
 \Big{(}1 + 2 \a (w_c+1) {\r_m \over M_{Pl}^2}  \Big{)}
 \Big{]}\label{L''sol}~.
\eea

A potential problem in the cosmology of the system would be, if
the
   denominator of \reef{A''sol} is equal to zero, \ie when $w_c=w_s^\pm$ with
\be w_s^\pm \equiv {-\left(1+4\a{ \r_m \over M_{Pl}^2} \right) \pm
\sqrt{64 \a^2{ \r_m^2 \over M_{Pl}^4} +32\a{ \r_m \over M_{Pl}^2}
+13  }  \over 12 \a{ \r_m \over M_{Pl}^2}}~. \label{singular} \ee
When this happens, the six-dimensional curvature invariant will
diverge close to the brane.
 Thus, after discussing the cosmological evolution of the brane world-volume, we should always check that
  it does not pass through a point which satisfies the above relation.

\def\thesubsection{\arabic{section}-\arabic{subsection}}

\subsection{Cosmological Evolution of an Isotropic
Brane-Universe}

\def\thesubsection{\arabic{section}-\arabic{subsection}.}

From the constraint relation \reef{isocon},  we can solve for
$w_c$, the allowed equation of state of the  matter on the brane.
It should satisfy the following equation \be w_c={-2 {\La_B \over
M_6^4} +{\r_m \over M_{Pl}^2} \left(1- {4 \over 3}\a{\r_m \over
M_{Pl}^2}\right) \over 3{\r_m \over M_{Pl}^2}\left(1+ {4 \over
3}\a{\r_m \over M_{Pl}^2}\right) } ~, \label{constrw} \ee with
$\r_m >0$ so that the Hubble parameter is real.

Before analyzing the system, let us note a first important
difference between the system of the pure four dimensional
  dynamics and the one with the extra constraint added because of the presence of the bulk.
   In the four dimensional system, a constant $w$ is allowed and its value is preserved during the evolution
    of the system. On the other hand, the evolution of the system with the extra constraint forbids
     any evolution with constant $w_c$. Indeed, by differentiating (\ref{constrw}) and using \reef{1rstfr}
       and the conservation equation $\dot{\r}_m+3(1+w_c)\r_m{\dot{a} \over a}=0$,
       we can find a differential equation for $w_c$
\be \dot{w}_c+3(1+w_c)\r_m {\de w_c \over \de \r_m}{\dot{a} \over
a}=0~. \label{wdifiso} \ee

Imposing a further condition of keeping $w_c$ constant, would
result to a constant $\r_m$, related to $w_c$ in a specific way,
 and by the conservation equation, to zero Hubble for $a$. Thus, an {\it a priori} fixing of $w_c$
  would result to an inconsistent system~\cite{Papantonopoulos:2005nw}.

To study the cosmological evolution, we look at the system \bea
\dot{H}&=&-{1 \over 2}(1+w_c){\r_m \over M_{Pl}^2}~,\\
\dot{\r}_m&=&-3(1+w_c)\r_m H~,\label{conserv1} \eea
 where
$H=\dot{a}/ a$ is the Hubble parameter.
 We will analyze the above system of the isotropic case for $\La_B=0$
  and $\La_B \neq 0$, because of the different features that arise in the two choices of this parameter.

  \def\thesubsubsection{\arabic{section}-\arabic{subsection}-\arabic{subsubsection}}

\subsubsection{Evolution of the System for $\La_B=0$}

\def\thesubsubsection{\arabic{section}-\arabic{subsection}-\arabic{subsubsection}.}

From the above dynamical system, taking into account
\reef{constrw}, we find that there is only one fixed point in the
evolution, the one with
 \be
(\r_m,H^2,w_c)=(0,0,1/3)~. \ee Linear perturbation around this
point reveals that it is an {\it attractor}.
 From \reef{wdifiso} and the conservation equation \reef{conserv1} we find
 for the Hubble parameter
\be {\dot{a} \over a}={2 \dot{w}_c \over
(1+3w_c)(1-3w_c)(1+w_c)}~. \ee The above equation can easily be
integrated and solved for $w_c$ (the $''-''$ sign in the solution
of $w_c$ is rejected because it gives imaginary  Hubble parameter)
\be w_c=-{1 \over 3} + {2 \over 3}~{a^4 \over \sqrt{3+a^8}}~. \ee
From this equation, we see that from any initial condition along
the line of tuning \reef{constrw},
 the expansion of the universe drives the equation of state to $w_c \to 1/3$, \ie radiation.
  We have also verified this by integrating the system numerically. During this cosmological evolution,
  it can never happen that $w_c=w_s^\pm$ (compare  \reef{singular} with \reef{constrw})
   and thus the whole system is regular.

\def\thesubsubsection{\arabic{section}-\arabic{subsection}-\arabic{subsubsection}}

\subsubsection{Evolution of the System for $\La_B \neq 0$}

\def\thesubsubsection{\arabic{section}-\arabic{subsection}-\arabic{subsubsection}.}

From the dynamical system, as written in the previous subsection,
taking into account \reef{constrw} with  $\La_B \neq 0$, we find
that there is a fixed point in the cosmological evolution \be
(\r_m,H^2,w_c)=\left(\r_f,{\r_f \over 3 M_{Pl}^2},-1\right)~, \ee
with ${\a\r_f \over M_{Pl}^2} = -{3 \over 4}+{3 \over 4}\sqrt{1+{4
\over 3}{\a\La_B \over M_6^4}}$. Since we should have $\r_m
>0$, this fixed point exists only for $\La_B>0$ and
corresponds to a de Sitter vacuum.  Linear perturbation around
this point reveals that it is an {\it attractor}. Thus, any matter
density on the brane eventually evolves to a state of vacuum
energy.

\begin{figure}[t!]

\begin{center}
\epsfig{file=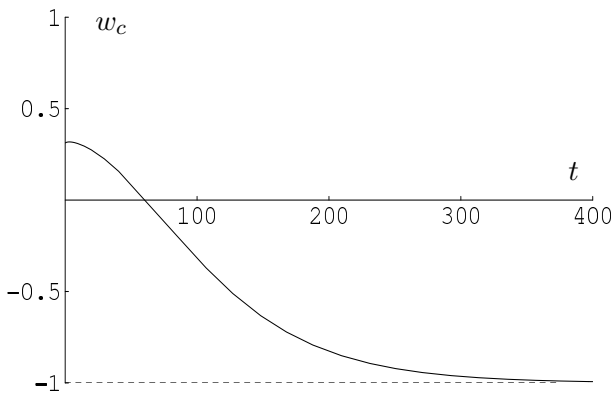,width=8cm,height=5cm}

\begin{picture}(.01,.01)(0,0)

\Text(100,95)[c]{$t$}

\Text(-75,150)[c]{$w_c$}

\end{picture}

\vskip-5mm

\caption{The  evolution of the equation of state $w_c$ for the
case of $\a \La_B / M_6^4 = 10^{-4}$ with initial conditions
$w_c=.31$ and ${\a \r_m \over M_{Pl}^2} =.02$.}

\label{smallLBiso}
\end{center}
\end{figure}

  A potentially interesting case, with a cosmological evolution
resembling that of our universe, would be the one where
\mbox{$0<\a\La_B / M_6^4 \ll 1$}. Then, the line of isotropic
tuning, as it is given by relation \reef{constrw}, has a maximum
close to $w_c \sim 1/3$. The evolution of an initial energy
density  larger than the one corresponding to that  maximum, will
evolve towards $w_c \sim 1/3$, pass from  $w_c=0$ and asymptote to
$w_c \to -1$ (see Fig.~\ref{smallLBiso} for an example). The
asymptotic  value of the  effective cosmological constant  at the
fixed point will be $\a {\La_{eff} \over M_{Pl}^2}  \sim \a {\La_B
\over M_6^4}\ll 1$ and should be rather small to match with
observations (when performing such a comparison,  it is reasonable
to assume  that all the dimensionful scales of the theory are
roughly of the same order, \ie  $\a^{-1/2} \sim M_{Pl} \sim M_6$).
The latter requirement of extremely small $\a\La_B / M_6^4$ is the
usual cosmological constant problem.  Although the standard
cosmological evolution is described by piecewise constant
equations of state with $w \sim 1/3,0,-1$, in the present theory
the  equation of state has always time dependence.

For $\La_B < 0$, there does not exist any fixed point. The
evolution of this system has a runaway behaviour and  flows to
$w_c \to \infty$ while $\r_m \to 0^+$.  During this evolution, the
singular point $w_c=w_s^+$ can be encountered only if
$\La_B/M_6^4 \lesssim -.326$ (equating the expressions
\reef{singular} and \reef{constrw} the energy density $\r_m$ can
be real and positive only for this range of $\La_B$). It is easily
verified that for these ``dangerous'' values of $\La_B$, it is
$w_s^+ > 1/3$. Therefore even in the case in which $\La_B$ takes
these values, the dangerous point $w_s^+$ is reached only if
initially  $w_c < w_s^+$,

We finally note that, for  $\La_B \neq 0$, there does not exist a
fixed point with  $(\r_m,H^2)=(0,0)$ because the equation of state
$w_c$ diverges at that point.

\def\thesection{\arabic{section}}

\section{The Unconstrained Anisotropic Case}

\def\thesection{\arabic{section}.}

In the previous section we found some interesting cosmological
evolutions of an isotropic brane-universe if $\La_B>0$. We should
keep in mind however, that in all cases studied, the energy
density on the brane is tuned to the equation of state
  in a specific way, which seems at first sight artificial.
    Therefore, it would be worth studying some cosmological evolution, in a codimension-2 brane-world model,
   in which this tuning is not required. If the system then evolves towards
    the previously studied line of isotropic tuning, we will conclude that
     this tuning is an attractor and thus not artificial.

To do so, we have to consider geometries which are not isotropic
 and in which the Riemann tensor cannot be expressed in terms of the Ricci tensor
  and the curvature scalar \cite{Papantonopoulos:2005ma}.
  Then,  the constraint equation \reef{rrcomp} will not give rise to a brane-bulk matter tuning,
   but rather to a dynamical equation for the anisotropy of the space.
    For this purpose, let us consider the following anisotropic ansatz where
     the metric functions depend only on the time $t$ and the radial coordinate $r$
      (\ie keeping the azimuthal symmetry)
\be
 ds^2= -N^2(t,r)dt^2 +\sum_{i=1}^3 A_i^2(t,r)(d x^i)^2 + dr^2 +
L^2(x,r) d\th^2~. \label{anisometric} \ee
 We can again use the gauge freedom to fix
$N(t,0)=1$, while we define $A_i(t,0)\equiv a_i(t)$.
 The singularity conditions dictate as before that $N'(t,0)=A'_i(t,0)=0$,
  while the second derivatives of these metric functions   are unconstrained.
   The most general anisotropic evolution scale factors with the above property can be written as
\bea
&&A_1(t,r)=a(t)b(t)c(t)+\xi_1(t)r^2 + \dots ~,\\
&&A_2(t,r)={a(t) \over b(t)}+\xi_2(t)r^2 + \dots~, \\
&&A_3(t,r)={a(t) \over c(t)}+\xi_3(t)r^2 + \dots \eea where
$a=(a_1a_2a_3)^{1/3}$  represents the ``mean'' scale factor and
$b$, $c$ represent two degrees of anisotropy.  To simplify further
the analysis of the system, we will choose $c=const.$
  (by a coordinate redefinition then we can always set $c=1$).
    The dynamics of this particular choice  can help us to understand the
   qualitative features of the general case.

The purely four-dimensional brane Einstein equations take the form
\bea
3{\dot{a}^2\over a^2} - {\dot{b}^2 \over b^2}&=&{\r_b +\La_4 \over M_{Pl}^2}~, \label{hubbleb}\\
2{\ddot{a} \over a}+4 {\dot{a}^2 \over a^2}&=&{\r_b-P_b +2\La_4 \over M_{Pl}^2}~, \label{accelb}\\
{\ddot{b} \over b}-{\dot{b}^2 \over b^2}+3{\dot{a}\dot{b} \over a
b}&=&0~, \label{bb} \eea
 while the equation which is
 coming from the extra dimensions is now dynamical,
providing a Hubble equation for $b$ \bea {\dot{b}^2 \over
b^2}&=&-{\r_b +\La_4 \over 4 M_{Pl}^2 }  \pm  {\sqrt{3} \over
16\a} \Big{\{} \sqrt{-16\a {\La_4\over M_{Pl}^2}
\Big{(}2 + \a {\La_4 \over M_{Pl}^2} \Big{)} } \nn \\
&+& \sqrt{{16 \a \La_B \over M_6^4} + {8\a \r_b \over M_{Pl}^2}
\Big{(}-1+2\a {\r_b
\over M_{Pl}^2} \Big{)} } \nn \\
&+& \sqrt{16\a {P_b \over M_{Pl}^2} \Big{(}{3 \over 2} +2\a {\r_b
+\La_4 \over M_{Pl}^2}\Big{)} } \Big{\}}~. \eea

We will now assume, as before, that the vacuum energy (tension)
 of the brane cancels the  contribution $\La_4$ induced my the deficit angle as in(\ref{relations}).
  After this simplification, the above equations read
\bea
3{\dot{a}^2\over a^2} - {\dot{b}^2 \over b^2}&=&{\r_m \over M_{Pl}^2}~, \label{Heq}\\
2{\ddot{a} \over a}+4 {\dot{a}^2 \over a^2}&=&(1-w){\r_m \over M_{Pl}^2}~, \label{addeq}\\
{\ddot{b} \over b}-{\dot{b}^2 \over b^2}+3{\dot{a}\dot{b} \over a
b}&=&0~, \label{bddeq} \eea while the equation coming from the
extra dimensions becomes \bea
 {\dot{b}^2 \over b^2}&=&-{\r_m \over 4
M_{Pl}^2 } + \sqrt{{3 \over 32\a}}\nn \\&~&\sqrt{2{\La_B \over
M_6^4}+ {\r_m \over  M_{Pl}^2 }\left[ (3w-1)+2(2w+1)\a {\r_m \over
M_{Pl}^2 }   \right]}\nn \\ &\equiv& f(\r_m,w)~. \label{fdef} \eea
  It is interesting to observe that the Hubble equation
\reef{Heq} for the ``mean'' scale factor $a$, after substitution
of \reef{fdef}, has apart from  the usual linear term in $\rho$
(of the conventional four-dimensional cosmology), additional
correction terms in $\rho$. This is similar to what happens also
to five-dimensional brane-world models \cite{5d}  and is  due to
the presence of extra dimensions. This modification happens only
in the anisotropic case.  In the pure isotropic case the
four-dimensional brane-universe feels the extra dimensions by only
adjusting its energy density to its equation of state,  but
without any modification in  the structure of the Friedmann
equation.

\def\thesubsection{\arabic{section}-\arabic{subsection}}

\subsection{Cosmological Evolution of an Anisotropic
Brane-Universe}

\def\thesubsection{\arabic{section}-\arabic{subsection}.}

Before analyzing the system, let us note again that in contrast to
pure four dimensional anisotropic dynamics,
 where  a constant $w$ is allowed, in the present case, where there is an extra dynamical
  equation because of the presence of the bulk,  $w$ has to evolve.
   Indeed, by differentiating (\ref{fdef})
    and then using the four dimensional equations of motion
     and the conservation equation $\dot{\r}_m+3(1+w)\r_m{\dot{a} \over a}=0$
we can find a differential equation for $w$ \be
 {\de f \over \de
w} \dot{w}+3\left[2f-(1+w)\r_m {\de f \over \de
\r_m}\right]{\dot{a} \over a}=0~. \label{wdifaniso} \ee
 Imposing a
further condition to keep $w$ constant, would result to a constant
$\r_m$ related to $w$ in a specific way and by the conservation
equation, to zero Hubble for $a$. Thus, an {\it a priori} fixing
of $w$ would result to an inconsistent
system~\cite{kofinas,Papantonopoulos:2005nw}.

To have real Hubble
 parameters  for  $a$ and $b$, $\r_m$ and $w$
have to lie in specific regions for which the following
inequalities are satisfied \be f>0~~~~~,~~~~~f+{\r_m \over
M_{Pl}^2}>0~. \label{allowedf} \ee
 Define the following boundaries
of the allowed regions \bea w_1&\equiv&{-2 {\La_B \over M_6^4}
+{\r_m \over M_{Pl}^2} \left(1- {4 \over 3}\a{\r_m \over
M_{Pl}^2}\right) \over 3{\r_m \over M_{Pl}^2}\left(1+ {4 \over
3}\a{\r_m \over M_{Pl}^2}\right) }~,
\label{w1}\\
w_2&\equiv&{-2 {\La_B \over M_6^4} +{\r_m \over M_{Pl}^2}
\left(1+4 \a{\r_m \over M_{Pl}^2}\right) \over 3{\r_m \over
M_{Pl}^2}\left(1+ {4 \over 3}\a{\r_m \over M_{Pl}^2}\right) }~.
\label{w2} \eea The quantity $w_1$ coincides with the line of
isotropic tuning $w_c$ given in \reef{constrw}. The inequalities
\reef{allowedf} can then be re-expressed as conditions for $\r_m$
and $w$:

\begin{itemize}

\item For $\r_m>0$, we should have $w>w_1$.

\item For $\r_m<0$, we distinguish two cases

\subitem - for $-{3 \over 4}<{\a \r_m \over M_{Pl}^2}<0$, we
should have $w<w_2$.

\subitem - for ${\a \r_m \over M_{Pl}^2}<-{3 \over 4}$, we should
have $w>w_2$.

\end{itemize}

To study the cosmological evolution, we look at the system \bea
\dot{H_a}&=&-{1 \over 2}(1+w){\r_m \over M_{Pl}^2}-f~,\label{Ha}\\
\dot{H_b}&=&-3H_a H_b~,\label{Hb}\\
\dot{\r}_m&=&-3(1+w)\r_m H_a~, \label{conserv} \eea
 where $H_a=\dot{a}/
 a$ and $H_b=\dot{b} / b$ are  the Hubble parameters
for $a$ and $b$ respectively. The third equation is the energy
conservation equation in which only the Hubble parameter $H_{a}$
for the ``mean''  scale factor $a$ appears.  Again we will analyze
the anisotropic case for $\La_B=0$ and $\La_B \neq 0$ and we will
compare the cosmological evolution with the cosmological evolution
of the tuned isotropic case.

\def\thesubsubsection{\arabic{section}-\arabic{subsection}-\arabic{subsubsection}}

\subsubsection{Evolution of the System with $\La_B=0$}

\def\thesubsubsection{\arabic{section}-\arabic{subsection}-\arabic{subsubsection}.}

There are three different regions in the $(w,\r_m)$ plane where
these inequalities are satisfied. This relative freedom to choose
the matter on the brane should
  be compared to the tuning that happens in the isotropic case.
  Relaxing the isotropy condition, the system can have initial
   conditions in a vast region of the parametric space.

From the dynamical system \reef{Ha}, \reef{Hb}, \reef{conserv} we
see that there is only one fixed point and that it is the same
with that of the isotropic evolution, \ie
 \be
(\r_m,H^2_a,H^2_b,w)=(0,0,0,1/3)~. \ee Linear perturbation around
this point reveals again that it is an {\it attractor}.

The presence of the previous attractor fixed point will drive the
system towards a final isotropic  state of radiation. The way in
which this fixed point is approached from an arbitrary initial
energy density, can tell us whether the line of isotropic tuning
is an attractor or not. If the anisotropy monotonically decreases
during the evolution, it means that the line of isotropic tuning
is an attractor.

In order to analyze the features of the anisotropic evolutions, we
proceed numerically. We solve the system of the two second order
equations \reef{addeq}, \reef{bddeq} and the two first order
equations \reef{wdifaniso}, \reef{conserv} for the four functions
$a$, $b$, $\r_m$, $w$. To understand how the anisotropy involves
we define  the mean anisotropy  by the following quantity \be
 A=\sqrt{\sum_{i=1}^3
{(\langle H \rangle -H_i)^2 \over 3 \langle H\rangle^2}}=\sqrt{{2
\over 3}}\left|{a \dot{b} \over \dot{a}b}\right|~, \ee
 where
$H_i=\dot{a}_i/ a_i$ (with $a_i$ defined after \reef{anisometric})
and $\langle H\rangle={1 \over 3}\sum_{i=1}^3 H_i=\dot{a}/ a$~.

Our analysis shows that, the relaxation of the tuning relation
between $w$ and $\r_m$, has as a consequence the appearance of new
branches of brane world evolution, while the system tends quickly
to the isotropic fixed point attractor. Furthermore, we observe
that the anisotropy in all cases decreases much more quickly than
in the four dimensional case with the same initial conditions but
without the extra dimensional constraint.

\def\thesubsubsection{\arabic{section}-\arabic{subsection}-\arabic{subsubsection}}

\subsubsection{Evolution of the System with $\La_B \neq 0$}

\def\thesubsubsection{\arabic{section}-\arabic{subsection}-\arabic{subsubsection}.}

Studying the asymptotics for $w_1$ and $w_2$ from \reef{w1},
\reef{w2} we find that for $\La_B \neq 0$, there are three
intervals of $\La_B$ with different shape of the allowed regions
in the $(\r_m,w)$ plane.

From the dynamical system \reef{Ha}, \reef{Hb}, \reef{conserv} we
see that there are two fixed points in general. The first one is
the same with that of the isotropic evolution, \ie
 \be
(\r_m,H^2_a,H^2_b,w)=\left(\r_f,{\r_f \over 3
M_{Pl}^2},0,-1\right)~, \ee
 with ${\a\r_f \over M_{Pl}^2} = -{3 \over 4}+{3 \over 4}\sqrt{1+{4
\over 3}{\a\La_B \over M_6^4}}$ and it exists only for $\La_B>0$.
Linear perturbation around this point reveals again that it is an
{\it attractor}.

The second fixed point that we find is
 \be
(\r_m,H^2_a,H^2_b,w)=\left(-\La_B{M_{Pl}^2 \over M_6^4},0,{\La_B
\over M_6^4},1\right)~. \ee and it exists only for $\La_B>0$.
Linear perturbation around this point reveals that it is a {\it
repeler}.

 Whenever  the previous attractor fixed point exists, the system
will be  driven towards a final isotropic  de Sitter state. The
way in which this fixed point is approached from an arbitrary
initial energy density, can tell us whether the line of isotropic
tuning is an attractor or not.  We analyzed  the anisotropic
system numerically and we found~\cite{Papantonopoulos:2005nw}
that, the relaxation of the tuning  between $w$ and $\r_m$, has as
a consequence the appearance of new branches of brane world
evolution, while for choice of the parameters the system tends to
the attractor fixed points, whenever they exist. For the cases
when $\La_B>0$ the lines of isotropic tuning are attractors, with
strength depending on the value of $\La_B$.

Furthermore, we observe that the anisotropy in all cases, apart
for $\La_B <0$, decreases much more quickly than in the four
dimensional case, with the same initial conditions, but without
the extra dimensional constraint. For $\La_B <0$, the anisotropy
increases and is larger than the one of the purely four
dimensional case.

 Let us now examine again  the interesting possibility of
\mbox{$0<\a\La_B / M_6^4 \ll 1$} with the transition between
$w=1/3$ to  $w=0$ and finally to $w=-1$. As it can be inferred
from the previous discussion, the system tracks the line of
isotropic tuning and evolves towards the attractor fixed point
with $w=-1$. However, due to the small value of $\a\La_B / M_6^4$,
the line of isotropic tuning is a {\it very weak attractor}. Most
of the evolution is rather anisotropic with $A \sim {\mathcal
O}(1)-{\mathcal O}(10^{-1})$  until  the fixed point is
approached, in which region it drops to zero.

This large anisotropy makes the cosmological evolution
phenomenologically problematic. In  order that the anisotropy is
acceptably small, the initial conditions  for the energy density
and the equation of state should be {\it fine tuned} to lie very
close to the line of isotropic tuning initially.

In conclusion, by analyzing the anisotropic dynamics of the system
we have seen that the lines of isotropic tuning are attractors for
$\La_B>0$, with $\La_B$-dependent attracting strength. The most
phenomenologically accepted evolutions with  $0<\a\La_B / M_6^4
\ll 1$ do not isotropise quickly enough and thus  need a fine
tuning in order to evolve with acceptable anisotropy.

\def\thesection{\arabic{section}}

\section{Six-Dimensional Supergravity Cosmological Models}

\def\thesection{\arabic{section}.}

There are also cosmological models in six dimensions coming from
supergravity theories~\cite{daemi2}. In some early works
cosmological solutions were found in D=6, N=2 supergravity
theories. These solutions describe vacuum state or radiation
four-dimensional universe in a Minkowski$\times S^{2}$
space~\cite{Maeda:1984gq}.

 Chaotic inflation was studied in an anolaly-free, gauged (1,0)
supergravity model in six dimensions in which the unique vacuum
state is Minkowski spacetime cross an internal $S^2$ which leaves
half of the six-dimensional supersymmetries unbroken~\cite{daemi}.
It was shown that inflationary dynamics consistent with the
cosmological constraints  can be realized provided that the radius
of the internal space $S^2$ satisfies a constraint. In this model,
the inflaton field $\phi$ originates from the complex scalar
fields in the D=6 scalar hyper-multiplet. The mass and the self
couplings of the scalar field are dictated by the D=6 Lagrangian.
The scalar potential has an absolute minimum at $\phi=0$ with no
undetermined moduli fields.

The model is based in a Lagrangian which have been constructed in
\cite{Nishino1,Nishino2}. It is chiral, hence potentially
inconsistent due to the presence of gauge, gravity and mixed
anomalies. In \cite{Randjbar-Daemi:wc}, an anomaly free model with
a gauge group $E_6\times E_7\times U(1)_R$ has been constructed
which involves a hyper-scalar multiplet transforming in a
912-dimensional pseudo-real representation of $E_7$. In a
convenient parameterization, the potential for the scalars take a
simple form
\begin{equation} \label{sixpot}
V= \frac{g_1^2}{\kappa^4}
e^{-\kappa\sigma}[(1+|\phi|^2)^2+\frac{g_7 ^2}{g_1 ^2 }(\phi
T^a\phi)^2]
\end{equation}
where $\kappa$ is the D=6 gravitational coupling and has a
dimension of square length. The $g_1$ and $g_7$ are the coupling
constants of the $U(1)_R$ and $ E_7$, respectively, and they have
the dimensions of a length, the $T^a$ are the Hermitian generators
of $E_7$ in its 912-dimensional representation. The most important
property of this potential is that it has a unique minimum at
$\phi=0$. Chaotic inflation is realized for the zero KK mode with
a potential (\ref{sixpot}) without the constant term.

\def\thesection{\arabic{section}}

\section{Conclusions}

\def\thesection{\arabic{section}.}

In this talk we discussed cosmological models in six dimensions.
First we presented a (4+1)-braneworld cosmological model in a
six-dimensional bulk. If $a(t)=b(t)$, with $a(t)$ the usual scale
factor of the three physical dimensions, and $b(t)$ the scale
factor of the extra fourth dimension, we found the generalized
Friedmann equation in six-dimensions of the Randall-Sundrum model
describing the cosmological evolution of a four-dimensional
brane-universe. If $a(t) \neq b(t)$ the four-dimensional universe
evolves with two scale factors. However, for an observer in the
moving brane, $a$ and $b$ are static depending only on the
coordinate on which the 4-brane is moving. Then, demanding to have
an effective Friedmann-like equation on the brane, we showed that
the motion of the 4-brane in the static bulk is constrained by
Darmois-Israel boundary conditions, resulting in a relation
 connecting $a$ and $b$ acting as a constraint of the brane motion.

 The way $a$ and $b$ are related depends on
the energy-matter content of the 4-brane. We then explored what
are the consequences of the presence of this constraint of motion
for the cosmological evolution of the brane-universe. We assumed
that the universe started higher-dimensional at the Planck scale
with all the dimensions at the Planck length, and subsequently an
anisotropy was developed between the three physical dimensions and
the extra-dimension. We then followed the evolution by making a
``phenomenological" analysis of how the two scale factors evolved
under various physical assumptions. In all considered cases, (A)dS
and Minkowski six-dimensional bulk, open, closed and flat
brane-universe and matter, radiation, cosmological constant and
dark energy dominated three-dimensional physical universe, we
found that dark energy is needed for the dynamical suppression and
subsequent freezing out of the extra fourth dimension.

We next discussed the gravitational dynamics of conical
codimension-2 branes
 of infinitesimal thickness. We have considered theories with codimension-2 branes which are augmented
    with an induced gravity term on the brane and a Gauss-Bonnet term in the
    bulk. We made the observation that the dynamics of the
these theories  is not exhausted
 by studying the ``boundary'' Einstein equation, which is exactly four-dimensional
  and bears no information of the internal space (modulo a cosmological constant contribution).
   In the case of a pure brane induced gravity term, the higher dimensional Einstein equations
    evaluated at the position of the brane,
    give a very precise and strong relation between the matter on the brane
     and the matter in the bulk in the vicinity of the brane.
         In other words, the bulk energy content is the primary factor for the cosmological evolution on the brane.
           Alternatively, for a static matter distribution on the brane to be possible,
            there should exist its bulk matter ``image''.

This strong relation, that we have noted, can be avoided with the
inclusion of a bulk Gauss-Bonnet term. An even more natural way to
achieve this is to relax the requirement of purely conical branes
and admit
  general brane solutions with an appropriate regularization (thickening of the brane),
   so that the singularities are smoothened.
    In view of the difficulties related to the Gauss-Bonnet term in the context of effective field theory,
     we pointed out that the thickening of the brane is the most physical
    direction that one should follow in order to discuss the dynamics of codimension-2 branes.

We then discussed the cosmological dynamics of a conical
codimension-2 braneworlds in a theory with a Gauss-Bonnet term in
the six-dimensional bulk and an induced gravity term on the
three-brane. For simplicity, we considered that the bulk matter
consists only of a cosmological constant $\La_B$ but the brane
matter is general and isotropic. We then analyzed in detail the
Einstein equations evaluated on the boundary.

We studied the system first for an isotropic metric ansatz. In the
pure induced gravity dynamics there is a tuning between the matter
allowed on the brane and in the bulk. This tuning, when the matter
in the bulk is only a cosmological constant, gives a precise
relation between the matter density on the brane and its equation
of state. If a Gauss-Bonnet term is added in the bulk, the
constraint equation giving the previous tuning is modified by the
addition of a Riemann squared term. However, since for isotropic
evolutions the  Riemann tensor can be expressed in terms of the
Ricci tensor and the scalar curvature, the conclusion about the
presence of the tuning remains the same as in the induced gravity
case. We found that
 if $\La_B>0$ the
system has a fixed point with equation of state $w=-1$ and
corresponds to a de Sitter vacuum. If $\La_B=0$, the system has a
fixed point with equation of state $w=1/3$. Both of the previous
fixed points are attractors. If $\La_B<0$ the system has no fixed
point and the evolutions exhibit a runaway behaviour to $w \to
+\infty$.

We then looked on the cosmological evolution of the system for a
particular anisotropy. If the system starts its evolution in the
region of parametric space which has as a boundary the line of
isotropic tuning, it tracks the latter line and isotropises
towards the attractor fixed point with $w=-1$ for $\La_B>0$, or
the one with $w=1/3$ for $\La_B=0$. In the two other regions the
system has a runaway behaviour $w \to\infty$ and does not
isotropise. For $\La_B<0$, the system has always a runaway
behaviour.  The important result of this analysis is that the line
of isotropic tuning, is an attractor. However, for values of
$\La_B$ which give acceptable cosmological evolutions, a fine
tuning is unavoidable because of the weak strength of the
above-mentioned attractor.

Six-dimensional supergravity models have a much richer structure.
There are many scalar fields with a scalar potential fixed by the
supergravity theory. There are also gauge fluxes which guarantee
the stability of the theory. Recently, they are six-dimensional
warped braneworld solutions in supergravity
teories~\cite{supergravity} and it would be interesing to find
also  codimension-1 or codimension-2 cosmological braneworld
solutions in six-dimensional supergravity theories.

\section*{Acknowlegements}
Most of the work reported in this talk was done in collaboration
with Bertha Cuadros-Melgar and Antonios Papazoglou and it is
supported by (EPEAEK II)-Pythagoras (co-funded by the European
Social Fund and National Resources).

\end{document}